\documentclass[10pt]{article}
\usepackage[english]{babel}
\usepackage[utf8x]{inputenc}
\usepackage[OT1]{fontenc}
\usepackage{amsfonts, amsmath, amsthm, amssymb}
\usepackage{graphicx}
\usepackage{listings}

\usepackage{multicol} 
\usepackage{amssymb, amsmath, amsbsy}
\usepackage{stackrel}

\usepackage{graphicx}
\usepackage{listings}
\usepackage{csquotes}
\usepackage[hidelinks]{hyperref}
\usepackage[nottoc,notlot,notlof]{tocbibind}

\usepackage{subcaption}
\usepackage[margin=1in]{geometry}
\usepackage{xcolor}
\usepackage{braket}

\title{Extrinsic Fluctuations in the p53 Cycle}

\author{
  \parbox{\linewidth}{ {\small
    Manuel Eduardo Hernández-García\textsuperscript{1,\dag},   Mariana Gómez-Schiavon\textsuperscript{2,3,\S} and Jorge Velázquez-Castro\textsuperscript{1,\ddag}}\\
    {\footnotesize
    \textsuperscript{1}Facultad de Ciencias Físico Matemáticas, Benemérita Universidad Autónoma de Puebla, Heroica Puebla de Zaragoza 72570, México. \textsuperscript{2}Laboratorio Internacional de Investigacion sobre el Genoma Humano, Universidad Nacional Autónoma de México, Santiago de Querétaro 76230, México.    \textsuperscript{3}Millennium Science Initiative Program, Millennium Institute for Integrative Biology (iBio), Chilean National Agency for Research and Development, Santiago 8331150, Chile.\\
    \textsuperscript{\dag} \texttt{manuel.hernandezgarcia@viep.com.mx}, \textsuperscript{\S} \texttt{mgschiavon@liigh.unam.mx}, \textsuperscript{\ddag}\texttt{jorge.velazquezcastro@correo.buap.mx}.
  }}}

\date{\today}
\begin{document}
\maketitle

\begin{abstract}
Fluctuations are inherent to biological systems, arising from the stochastic nature of molecular interactions, and influence various aspects of system behavior, stability, and robustness. These fluctuations can be categorized as intrinsic, stemming from the system's inherent structure and dynamics, and extrinsic, arising from external factors, such as temperature variations. Understanding the interplay between these fluctuations is crucial for obtaining a comprehensive understanding of biological phenomena. However, studying these effects poses significant computational challenges. In this study, we used an underexplored methodology to analyze the effect of extrinsic fluctuations in stochastic systems using ordinary differential equations instead of solving the Master Equation with stochastic parameters. By incorporating temperature fluctuations into reaction rates, we explored the impact of extrinsic factors on system dynamics. We constructed a master equation and calculated the equations for the dynamics of the first two moments, offering computational efficiency compared with directly solving the chemical master equation. We applied this approach to analyze a biological oscillator, focusing on the p53 model and its response to temperature-induced extrinsic fluctuations. Our findings underscore the impact of extrinsic fluctuations on the nature of oscillations in biological systems, with alterations in oscillatory behavior depending on the characteristics of extrinsic fluctuations. We observed an increased oscillation amplitude and frequency of the p53 concentration cycle. This study provides valuable insights into the effects of extrinsic fluctuations on biological oscillations and highlights the importance of considering them in more complex systems to prevent unwanted scenarios related to health issues.\\

\end{abstract}

\newpage
{
\tableofcontents }

\newpage

\section{Introduction}
Fluctuations are inherent to biological systems, arising from the stochastic nature of molecular interactions, and influence various aspects of system behavior, stability, and robustness. The fluctuations play an even more pivotal role, necessitating thorough consideration in comprehensive analyses \cite{Samo, Alon, Vechio, Gar}. These fluctuations can be broadly categorized as intrinsic and extrinsic, with each contributing distinctively to system dynamics.

Intrinsic fluctuations stem from the inherent structure and dynamics of the system, reflecting the probabilistic nature of molecular interactions within biological networks \cite{Jong, Gomez, Manu, Simpson}. These fluctuations manifest as variabilities in molecular concentrations influencing the emergence of spontaneous oscillations, bistability, and other behaviors observed in biological systems \cite{Samo, Alon, Vechio, Gar}. 

In contrast, extrinsic fluctuations arise from external factors, such as variations in temperature, pH, or nutrient availability, which can significantly affect system dynamics \cite{Hilfi}. Although intrinsic fluctuations have received considerable attention in previous research, the importance of extrinsic fluctuations has garnered increasing recognition. Understanding the interplay between intrinsic and extrinsic fluctuations is crucial for obtaining a comprehensive understanding of biological phenomena, because both types of fluctuations can coalesce to shape system behavior in experimental setups and natural environments \cite{Arriaga}.

Despite the significance of the fluctuations, studying their effects poses significant computational challenges. Existing algorithms and analytical methods may struggle to capture the complex interactions between intrinsic and extrinsic fluctuations accurately. Moreover, computationally intensive approaches are often required to discern the effects of extrinsic fluctuations, further complicating analyses \cite{Vastola, Vol, Zech, Ham, Toni, Cara, Swain}.

Therefore, alternative analytical approaches are required to address these challenges.  We adopted an approach based on the calculation of the master equation with discrete quantities, considering both the molecules involved in the reactions and the parameters that influence biochemical reaction rates. This implies that each molecule and parameter is treated as stochastic variables in the proposed model, extending prior methodologies to encompass a broader range of scenarios \cite{ Vastola, Ham}. By approximating the master equation, we derive a set of differential equations that model mesoscopic system dynamics \cite{ Gomez, Manu}; in particular, in scenarios involving small fluctuations and/or first-order reactions, this approximation becomes exact. This approach offers computational efficiency compared with traditional stochastic simulations, such as the Gillespie algorithm \cite{Gillespie}.

The generation of biological oscillators, such as circadian rhythms, is a complex process involving the expression and cyclic regulation of specific genes, as well as the feedback mechanism between proteins \cite{Nova}. These mechanisms govern diverse gene regulatory networks and other biological processes \cite{Nova, Roe, Vecchio, Briat}. Here, we focused on assessing the robustness of biological oscillators under extrinsic fluctuations with specific attention to the p53 system. As a critical regulator of cellular stress responses and the maintenance of genomic integrity, the p53 system's oscillatory behavior enhances response precision \cite{Vou, Brew, Zato, Batch}.  Examining the influence of temperature fluctuations, which are known to significantly affect biochemical reactions \cite{Ko, Busse, Dere, Pil, Stri}, we must consider that such fluctuations are present in the p53 system. Therefore, it is essential to assess how the system functions in the presence of both intrinsic and extrinsic fluctuations. In this work, we aim to determine the sensitivity of the system dynamics to fluctuations. With this, we aim to gain a deeper understanding of the effects of fluctuations on the biological oscillations that occur in living organisms. \\

The remainder of this paper is organized as follows. Section \ref{section2} outlines the derivation of the master equation considering extrinsic fluctuations. In Section \ref{section 3}, we employ a deterministic approximation to derive differential equations that describe the system dynamics in terms of concentrations and their covariances. Section \ref{section4} focuses on modeling the p53 system and evaluating the effects of temperature fluctuations on its dynamics. Finally, in Section \ref{section5} the results and conclusions are presented.

\section{Master Chemical Equation with Extrinsic Components} \label{section2}
The master chemical equation of a biochemical system is responsible for modeling processes involving interactions between various biochemical species and their fluctuations. However, in some cases, these interactions are influenced by the surrounding environment (e.g., temperature and pressure) and their fluctuations, which are referred to as extrinsic fluctuations. 

Initially, it is essential to consider the chemical equation for a system that remains uninfluenced by external elements \cite{Gar}. This equation, referred to as the master chemical equation, describes chemical reactions that occur within a system. To achieve this, we assume the presence of $N$ species, $\mathcal{S}_l$ ($l$ $\in$ \{$1,2,...,N$\}), and $M$ reactions $\mathcal{R}_j$ ($j$ $\in$ \{$1,2,...,M$\}) in which the species are transformed as follows,
\begin{align}
    \mathcal{R}_j : \sum_{l=1}^{N} \alpha_{jl} \mathcal{S}_l \stackbin[]{k_{j}}{\rightarrow} \sum_{l=1}^{N} \beta_{jl} \mathcal{S}_l. \label{bd1}
\end{align}
Here, coefficients $\alpha_{jl}$ and $\beta_{jl}$ are positive integers and $k_j$ are the reaction constants (or parameters). We assume that the reactions follow mass action kinetics. These equations allow for the determination of the stoichiometric matrix of the system $\Gamma_{lj}= \beta_{jl} -  \alpha_{jl}$. Through the collisions (or interactions) of the different elements they are transformed, so the propensity rates are given as follows
\begin{align}
    {t_j(\mathbf{S})}&= k_{j} \prod_{l}^N \frac{S_l !}{\Omega ^{\alpha_{jl}}(S_l- \alpha_{jl} )!}, \label{bd2}
\end{align}
(the index $j$ is the same as that of reaction $\mathcal{R}_j$), which are the transition probabilities between different states of the system. $S_l$ corresponds to the number of molecules of the chemical species $\mathcal{S}_l$ and $\mathbf{S}=$ ($S_1,S_2,...,S_N$),  $\Omega$ is related to the size of the system and has units of molecules between moles. Based on this background, the following equation can be derived,

{\small
\begin{align}
    \frac{d P(\mathbf{S},t)}{dt} = \Omega \sum_j^{M} \left( t_{j}(\mathbf{S}-\Gamma_{j}) P(\mathbf{S}-\Gamma_{j},t) - t_{j}(\mathbf{S}) P(\mathbf{S},t) \right), \label{4}
\end{align}}
this is the chemical master equation, which describes the evolution of the probability distribution of states in a system. We can make the dependence of $P(\mathbf{S},t)$ on the reaction constants $\mathbf{K}=(K_{1},K_{2},...,K_{M})$ explicit by writing (\ref{4}) as
\begin{equation}
    \frac{d P(\mathbf{S},t|\mathbf{K})}{dt} = \Omega \sum_j^M \left( T_{j}(\mathbf{S}-\Gamma_{j},K_{j}) P(\mathbf{S}-\Gamma_{j},t|\mathbf{K}) - T_{j}(\mathbf{S},K_{j}) P(\mathbf{S},t|\mathbf{K}) \right),
\end{equation}
where 
\begin{align}
{T_j(\mathbf{S}, {K}_j)}&= K_{j} \prod_{l}^N \frac{S_l !}{\Omega ^{\alpha_{jl}}(S_l- \alpha_{jl} )!}. \label{7}
\end{align}
Furthermore, reaction constants ${K}_j$ can be time-dependent,  $K_{j}=K_{j}(t)$, and follow stochastic dynamics. We considered these fluctuations by following a procedure similar to that in \cite{Vastola, Ham}.

\subsection{Fluctuations in Reaction Constants}
Fluctuations in reaction rates can be driven by extrinsic factors such as temperature or density fluctuations because the system is not isolated, and thus in thermodynamic contact with the environment.
Let's consider $M$ time-dependent reaction rates, 
\begin{align}
        K_{j}(t) =  k_{j}(t) + \eta_{k_j}(t), 
\end{align}
($j \in$ ($1,2,...,M$)). These are expressed as the mean dynamics $k_{j}$ and  $\eta_{k_j}$ is a stochastic term representing the fluctuations around this mean (for more details, see Appendix \ref{A}). Thus, we have
     
\begin{align}
          k_{j}(t) & = \braket{K_{j}(t)}, \nonumber \\
         C^2_{l,j} &=\braket{({K}_{l}(t)- k_{l}(t)) ({K}_{j}(t)- k_{j}(t) )}=\braket{\eta_{k_l}\eta_{k_j}}.
 \end{align}
 Additionally, to describe the dynamics of the reaction constants, we use the following Fokker-Planck equation
 \begin{equation}
         \frac{\partial p(\mathbf{K}, t)}{\partial t} = -\sum_{j}^M \frac{\partial}{\partial K_j}g_{j}(\mathbf{K},t)p(\mathbf{K}, t) + \frac{1}{2}\sum_{j}^M \frac{\partial^{2}}{\partial K_{j} K_l} G_{lj}(\mathbf{K},t)p(\mathbf{K}, t),
\end{equation}
where $p(\mathbf{K}, t)$ describes the probability distribution of $\mathbf{K}$, $g_{j}$ are the drift coefficients and $G_{ji}$ are the diffusion coefficients. As $\mathbf{K}$ is now a stochastic variable, we can now define the joint probability density
     \[ P(\mathbf{S},\mathbf{K},t)=P(\mathbf{S},t|\mathbf{K})p(\mathbf{K},t), \]
     then 
     \begin{align}
         \frac{\partial P(\mathbf{S},\mathbf{K},t)}{\partial t} = \Omega \sum_j^M \left( T_{j}(\mathbf{S}-\Gamma_{j}, {K}_j(t)) P(\mathbf{S}-\Gamma_{j},\mathbf{K},t) - T_{j}(\mathbf{S}, {K}_j(t)) P(\mathbf{S},\mathbf{K},t) \right)  \nonumber  \\
          -\sum_{j}^M \frac{\partial}{\partial K_j}g_{j}(\mathbf{K},t)P(\mathbf{S},\mathbf{K}, t) + \frac{1}{2}\sum_{j}^M \frac{\partial^{2}}{\partial K_{j} K_l} G_{lj}(\mathbf{K},t)P(\mathbf{S},\mathbf{K}, t)
    \label{8}
     \end{align}
we obtain a master chemical equation that includes the Fokker-Planck equation, which is equivalent to that derived in \cite{Vastola}.  Equation (\ref{8}) models the distribution of a system's states under the influence of external stimuli, allowing for fluctuations (extrinsic fluctuations).  This approach enhances the precision of biological system studies by aligning them more closely with the fluctuating nature of the systems.  

\section{Approximation with Extrinsic Fluctuations} \label{section 3}
A system with intrinsic and extrinsic fluctuations can be analyzed directly using the master equation (\ref{8}),  for this the Gillespie algorithm is used \cite{Gillespie} and/or solving the Fokker-Planck equation.  However, it is more practical and computationally efficient to use an approximation than to use the master equation directly. 

For example, the approximation presented in \cite{Gomez, Manu} can be used. This approximation transforms the master equation into a set of differential equations of moments, specifically the means and covariances, and assumes that higher-order moments are zero because they do not significantly contribute to system dynamics \cite{Gomez}. However, we propose another interpretation in which the fluctuations of the variables are sufficiently small; if we have a function of the variables of the system, a second-order expansion around their mean values is sufficient to capture the effects of the fluctuations. In this case, averaging was performed for a function of the variables $S$ and $K$; therefore, we used $\mathbf{X}=$\{$\mathbf{S}, \mathbf{K}$\}, and $\hat{X}_j=\braket{X_j}$ ($X_j$ is the $j$ element of the vector $\mathbf{X}$). To obtain the differential equations, we will use the following approach,

{\small
\begin{align}
    \braket{f(\mathbf{X})}_{\mathbf{X}} \approx & \left\langle f(\braket{\mathbf{X}}) + \sum_{j_1} (\braket{X_{j_1}}-X_{j_1})\frac{\partial f(\mathbf{\hat{X}})}{\partial \hat{X}_{j_1}} + \sum_{j_1} \sum_{j_2} \frac{1}{2} (\braket{X_{j_1}}-X_{j_1})(\braket{X_{j_2}}-X_{j_2}) \frac{\partial^2 f(\mathbf{\hat{X}})}{\partial \hat{X}_{j_1} \partial \hat{X}_{j_2}}  \right\rangle_{\mathbf{X}} \nonumber \\
    =& f(\mathbf{\hat{X}}) + \sum_{j_1} \sum_{j_2} \frac{C({X}_{j_1},{X}_{j_2})}{2} \frac{\partial^2 f(\mathbf{\hat{X}})}{\partial \hat{X}_{j_1} \partial \hat{X}_{j_2}}.\label{11}
\end{align}}


\noindent Where the $\hat{X}_i$'s are related with deterministic quantities.  Note that the average is taken for all system variables, whereas $\mathbf{C}_{j_1,j_2}$ denotes the covariance between the different variables. Note that if $f(\mathbf{X})$ is a second-order polynomial, then the approximation becomes exact. 

\noindent This expansion around the mean value is of great usefulness when applied in conjunction with the master equation. For example, we can use (\ref{8}) to obtain an expression for the evolution of $s_{l} = \frac{ \braket{S_l}}{\Omega}$ and $k_j= \braket{K_j}$  and then we use the expansion (\ref{11}) to obtain the following equations,

{\small
\begin{align}
     \frac{d s_l}{dt}=& \sum_{j}^{M} \Gamma_{lj}\left( R_j^{D}(\mathbf{s})  + \sum_{l_1}^{N} \sum_{l_2}^{N}  \frac{\sigma^2_{l_1,l_2}}{2}  \frac{\partial^2 R_j^{D}(\mathbf{s})}{\partial {s_{l_1}}\partial {s_{l_2}}}  +  \underline{ \sum_{l_1}^{N}  {C^1_{l_1,j}} \frac{\partial R_j^{D}(\mathbf{s})}{\partial {s_{l_1}}}  } \right), \label{10}\\
    \frac{d {k_j}}{dt}=& g_j(\mathbf{k},t) +  \sum_{j_1}^{M} \sum_{j_2}^{M}  \frac{C^2_{j_1,j_2}}{2}  \frac{\partial^2 g_j(\mathbf{k},t)}{\partial {k_{j_1}}\partial {k_{j_2}}}     ,    \label{13}
\end{align}}

where $l=$ \{$1,2,..., N$\}, $j=$ \{$1,2,...,M$\}, and $R_j^{D}(\mathbf{s})= k_j \prod_l^N s_l^{\alpha_{jl}}$ is the deterministic reaction rate assuming mass action, and the covariances between variables  $\sigma^2_{l_1,l_2}= \frac{\braket{(S_{l_1}-s_{l_1})(S_{l_2}-s_{l_2})}}{\Omega^2}$, $C^2_{{j_1},{j_2}}= \braket{(K_{j_1}-k_{j_1})(K_{j_2}-k_{j_2})}$, $C^1_{l_1,{j}}= \frac{\braket{(S_{l_1}-s_{l_1})(K_{j}-k_{j})}}{\Omega}$, $\mathbf{s}=$ ($s_1,s_2,...,s_N$), $\mathbf{k}=$ ($k_1,k_2,...,k_M$). In contrast with the case without extrinsic fluctuations or constant reaction rates $k_{j}$ \cite{Manu}, the evolution of the mean concentrations $s_{j}$ (in what follows we will refer only how concentration) is influenced by the correlations between the $K_{j}$'s and the concentrations themselves as can be seen from the underlined terms of equation (\ref{10}). 
 
With a similar procedure, we can now calculate a set of equations for the evolution of the covariances between variables $s_l$ of the system,  
 
{\footnotesize
\begin{align}
    \frac{d \sigma^2_{l_1,l_2}}{dt} &=    \sum_j^{M}  \left( \frac{\Gamma_{l_1 j}\Gamma_{l_2j}}{ \Omega} \left( R_j^{D}(\mathbf{s})  + \sum_{l_3}^{N} \sum_{l_4}^{N}  \frac{\sigma^2_{l_3,l_4}}{2}  \frac{\partial^2 R_j^{D}(\mathbf{s})}{\partial {s_{1_3}}\partial {s_{l_4}}}  +  \underline{ \sum_{l_3}^{N}  \frac{C^1_{l_3,j}}{k_j}  \frac{\partial R_j^{D}(\mathbf{s})}{\partial {s_{l_3}}}  } \right) \right.  \nonumber \\
    &  + \sum_{l_3}^{N}   \left( {\sigma^2_{l_1,l_3}} \Gamma_{l_2 j} \frac{\partial R_j^{D}(\mathbf{s})}{\partial {s_{l_3}}} +{\sigma^2_{l_3,l_2}} \Gamma_{l_1 j} \frac{\partial R_j^{D}(\mathbf{s})}{\partial{s_{l_3}}}\right)  + \left. \underline{ \Gamma_{l_1 j} \left(\frac{C^1_{l_2, j}}{k_j}   R_j^{D}(\mathbf{s}) \right)   + \Gamma_{l_2 j} \left(\frac{C^1_{l_1, j}}{k_j}   R_j^{D}(\mathbf{s}) \right)} \right) , \label{15}
\end{align}}

\noindent where $l_1, l_2=$ \{$1,2,..., N$\}. To highlight the effect of extrinsic fluctuations we have underlined the terms that are contributing to the dynamics of the correlation due to the fluctuating environment described by the fluctuation rate constants. This kind of effect in the dynamics of the correlations was already observed in \cite{Hilfi, Eloe}.

To solve the previous differential equations we need to add equations that describe the evolution of $C^1_{l,j}$ and $C^2_{i,j}$, these are obtained in a similar way that Equation (\ref{15})

{\small
\begin{align}
    \frac{d C^1_{l,j}}{d t} =& \sum_i^{M} \Gamma_{{l}i}\left(    \frac{C^2_{i,j}}{k_i}   R_i^{D}(\mathbf{s}) +  \sum_{l_1}^{N}  C^1_{l_1,j} \frac{\partial R_i^{D}(\mathbf{s})}{\partial{s_{l_1}}} \right) + \sum_{j_1}^{M} C^1_{l,j_1} \frac{\partial g_j(\mathbf{k},t)}{\partial k_{j_1}} , \label{16} \\
    \frac{d C^2_{i,j} }{dt}=&  C^2_{i,j} \left( \frac{\partial g_i(\mathbf{k},t) }{ \partial k_{i}} + \frac{\partial g_j(\mathbf{k},t) }{ \partial k_{j}} \right)+ \left( G_{ij}(\mathbf{k},t) + \sum_{j_1}^{M} \sum_{j_2}^{M} \frac{C^2_{j_1,j_2}}{2} \frac{\partial^2 G_{ij}(\mathbf{k}, t) }{ \partial k_{j_1} k_{j_2}} \right) .     \label{17}
\end{align}}

Where $i=$ \{$1,2,...,M$\}. Solving the set of differential equations (\ref{10})–(\ref{17}) allows us to describe the dynamics of the mean and variance of species concentration in the presence of both intrinsic and extrinsic fluctuations. Using equations (\ref{10})–(\ref{17}) for a particular case, as presented in \cite{Vastola}, we can reproduce their results (see Appendix \ref{C}) while also providing new insights and extending their work by applying it to another type of extrinsic fluctuations. 

\begin{figure}[h!t]
\centering
\includegraphics[width=0.4\textwidth]{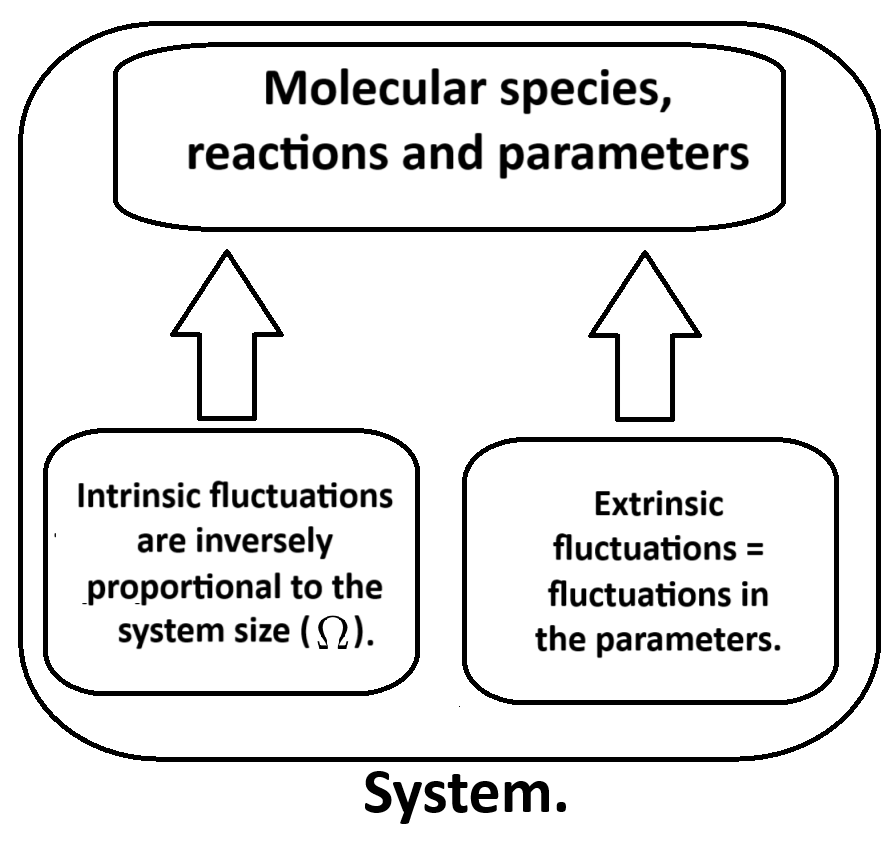}
\caption{\textbf{Interplay of intrinsic and extrinsic fluctuations over a system.} This figure illustrates how fluctuations affect the molecular species, reactions, and parameters. The magnitude of intrinsic fluctuations is primarily determined by the inverse of the system size ($\Omega$). Conversely, extrinsic fluctuations arise exclusively from fluctuations in parameters.}
\label{fig_diagrama}
\end{figure}

It is worth noting that, in contrast to intrinsic fluctuations, for which the magnitude decreases with system size $\Omega$, the magnitude of extrinsic fluctuations depends only on external factors; thus, for large systems, when intrinsic fluctuations become negligible, the stochastic behavior of the species concentrations $s_{j}$ is solely due to extrinsic fluctuations.  

In the following section, we use differential equations (\ref{10})–(\ref{17}) to analyze a particular system, representing a model of p53.

\section{Model of p53} \label{section4}
The protein p53, which is crucial for cancer prevention, plays a vital role in maintaining genomic integrity and regulating cellular fate under normal conditions. However, its dysfunction, which is common in many cancers, can have serious consequences.  When p53 fails to detect and repair DNA damage, it promotes the accumulation of mutations and tumor progression by allowing the survival of damaged cells \cite{Vou, Brew, Zato, Batch, Hun}. Oscillations in p53 activity are relevant because they potentially fine-tune cellular processes such as the cell cycle, suggesting a significant function in coordinating the cellular response to stress and DNA damage. This enables a more adaptive and efficient response to changing conditions \cite{Zato}. Characterizing these oscillations is crucial for understanding cancer suppression mechanisms and developing new therapies. Therefore, we analyzed the effect of body temperature fluctuations on the period and amplitude of these oscillations.

\begin{figure}[h!t]
\centering
\includegraphics[width=0.5\textwidth]{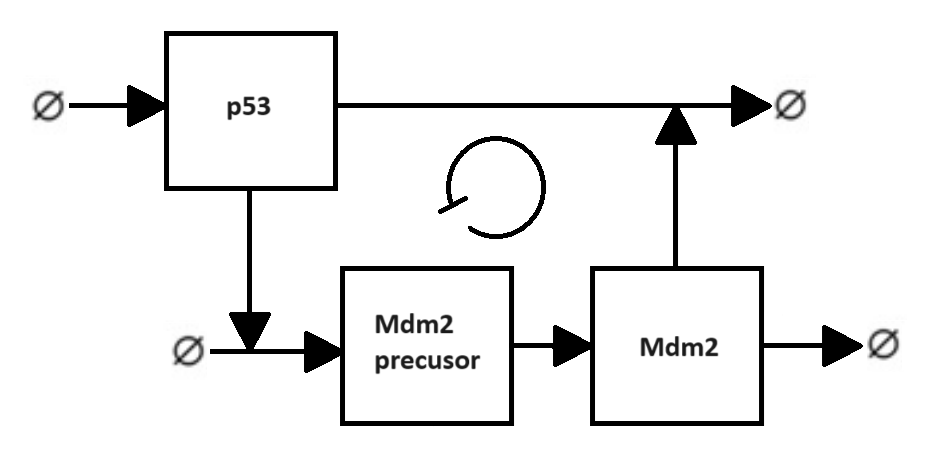}
\caption{\textbf{Model of p53.} p53 induces the synthesis of Mdm2 precursor, which eventually leads to the production of Mdm2. In turn, Mdm2 promotes the degradation of p53, creating a negative feedback loop.}
\label{fig0}
\end{figure}

We adopted a model presented in \cite{Zato} which describes the dynamics of the biochemical species concentrations. This particular model is a simplified representation of p53 dynamics, as it does not consider the presence of other reactions \cite{Batch}. We chose this model because its dynamic is not affected by $\Omega$, a factor that influences some other models (see Appendix \ref{D}). To induce oscillations in the system, we consider the parameters and initial conditions provided in \cite{Zato}. The model, depicted in Figure \ref{fig0}, which contains three types of molecules: p53, Mdm2-precursor, and Mdm2. p53 promotes the synthesis of Mdm2-precursor, which in turn promotes the synthesis of Mdm2. However, Mdm2 promotes the degradation of p53. This interaction is modeled through a Hill-type function, Mdm2 binds to p53, leading to its degradation. Additionally, both p53 and Mdm2 are subject to degradation. 

The system is described by the following reactions,
\begin{table}[h!]
    \centering
    \begin{tabular}{|c|c|}
        $\emptyset  \stackbin{k_1}{\longrightarrow} P53$ & $P53  \stackbin{k_2}{\longrightarrow} \emptyset$ \\
        $P53 + \text{Mdm2}  \stackbin{k_3^*}{\longrightarrow} \text{Mdm2}$ & $P53  \stackbin{k_4}{\longrightarrow} P53 + \text{Mdm2 precursor}$ \\
        $\text{Mdm2 precursor}  \stackbin{k_5}{\longrightarrow} \text{Mdm2}$ & $\text{Mdm2} \stackbin{k_6}{\longrightarrow} \emptyset$ \\
    \end{tabular}
\end{table}

For this system, we have the next stoichiometric matrix
{\small
\begin{align}
 \Gamma_{ij}&= \begin{pmatrix}
		1& -1 & -1& 0 & 0 &0 \\
		0& 0 & 0 & 1 & -1 &0 \\
        0& 0 & 0 & 0 & 1  &-1
	\end{pmatrix}. \nonumber
\end{align}}
Here, $x_1$, $x_2$, and $x_3$ represent the concentrations of p53, Mdm2-precursor, and Mdm2, respectively. $k_3^*={k_3} \left( \frac{1}{A_1 + x_1} \right)$, $A_1$ is the p53 threshold for degradation by Mdm2.  $k_1$ is the p53 synthesis rate, $k_2$ is the p53 degradation rate, $k_3$ is the saturated p53 degradation rate,  $k_4$ is the p53-dependent Mdm2 production rate, $k_5$ is the Mdm2 maturation rate and $k_6$ is the Mdm2 degradation rate. The reaction rates of the system are

\begin{align}
   R_1^{D}&= k_1 , &
    R_2^{D}= k_2 x_1, \nonumber \\
   R_3^{D}&= {k_3} x_3 x_1 \left( \frac{1}{A_1 + x_1} \right) , &
    R_4^{D}= k_4 x_1, \nonumber \\
    R_5^{D}&= k_5 x_2  , &
    R_6^{D}= k_6 x_3. \label{18}
\end{align}

A Hill-type function appears in the reaction rates (\ref{18}) associated with the degradation of p53, when such functions appear, the result in \cite{Manu} is used to aid in modeling stochastic systems, because in this work is derived a stochastic Hill function,  that considers and incorporates the effects of the fluctuations directly in the stochastic Hill function. 

\subsection{Case I: Stationary extrinsic fluctuations}

In this case, it is assumed that the $k_j$'s and the magnitude of their fluctuations do not change over time. In other words, we have stationary extrinsic fluctuations. Using the Arrhenius equation \cite{Atkin} it is possible to relate $C^2_{i,j}$ with the variance of the temperature  $\sigma^{2}_{T}$  as follows 
\begin{align} 
    C^2_{i,j}= \left(\frac{k_i}{10}\right)\left(\frac{k_j}{10}\right) \sigma^2_T , \label{20}
\end{align}
for $i,j=$ \{$1,2,3,4,5,6$\} (see Appendix \ref{A} for further details). 

\begin{figure}[h!t]
\begin{subfigure}{\linewidth}
\includegraphics[width=0.5\textwidth]{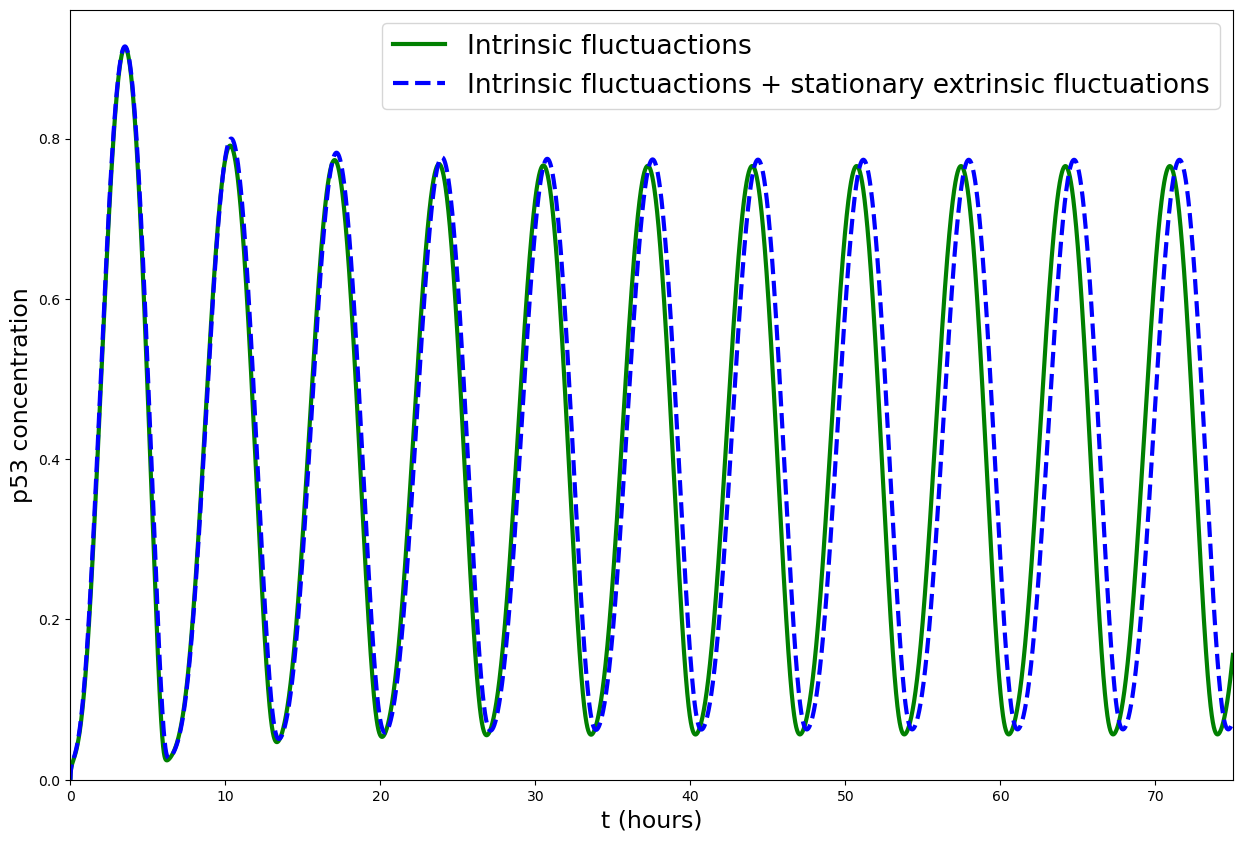}\hfill
\includegraphics[width=0.5\textwidth]{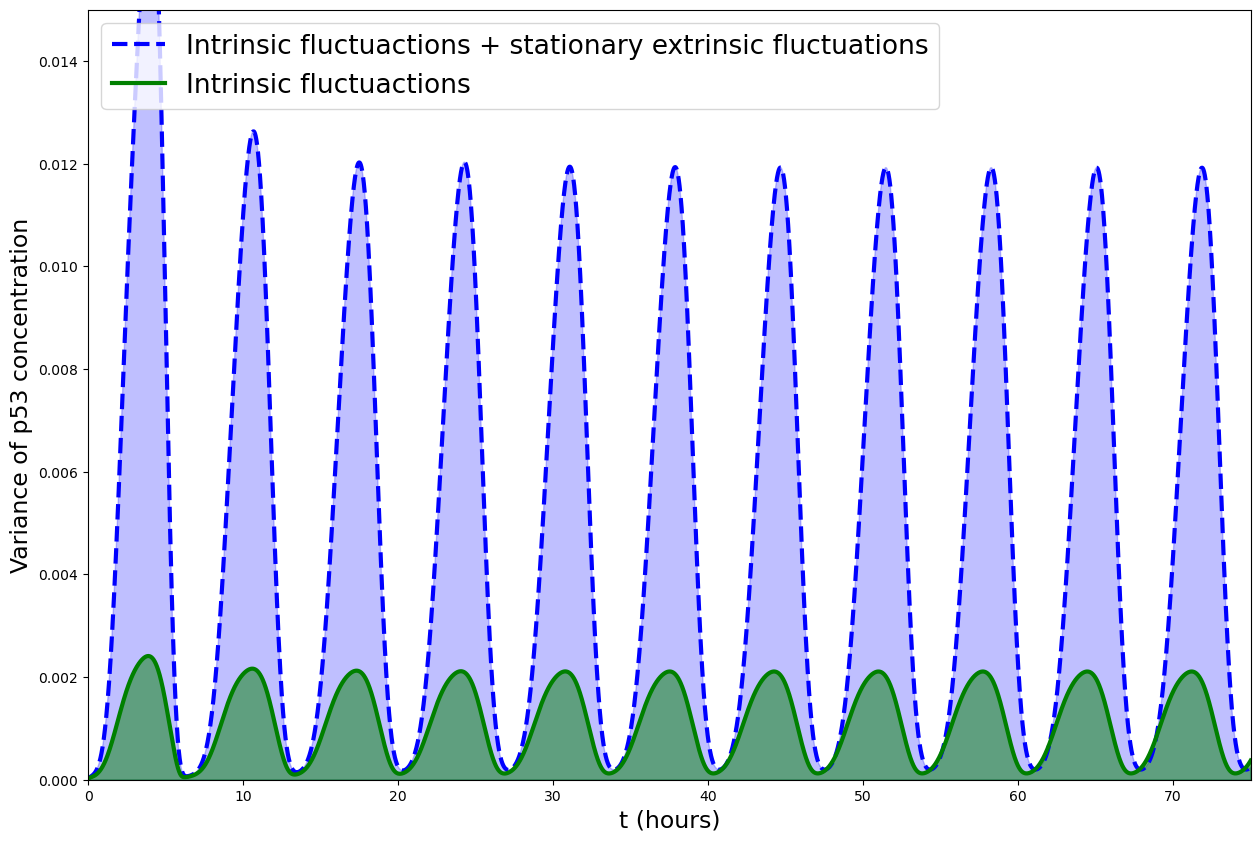}
\end{subfigure}
\caption{ {\textbf{Dynamics of p53 concentration and its variance with stationary extrinsic fluctuations.} Left: Concentration dynamics of the p53 under two conditions: only intrinsic fluctuations (green line), and both intrinsic and extrinsic fluctuations (dashed blue line). When considering both types of fluctuations, we observed an increase in the amplitude and phase shift compared to when only intrinsic fluctuations were present. Right: Variance of p53 concentration. Extrinsic fluctuations alongside intrinsic fluctuations, result in an amplified amplitude and phase shift compared to the scenario with only intrinsic fluctuations. The system was modeled with the parameters and initial conditions listed in Table \ref{tabla1}. In the case of extrinsic fluctuations, Table \ref{tabla2} was used with $\sigma^2_T=1$.}}
\label{fig1}
\end{figure}

We solved the system numerically, the results are shown in Figure \ref{fig1}. The left panel shows the temporal evolution of p53 concentration when only intrinsic fluctuations are considered and when intrinsic and stationary extrinsic fluctuations are taken into account. The right panel shows the temporal evolution of the p53 variance in both cases. 

In \cite{Lakatos} a similar model was analyzed, but the oscillations dampened and the variance increased indefinitely; however, in Figure \ref{fig1}, we can observe that in both cases (only intrinsic fluctuations and intrinsic and extrinsic fluctuations), the oscillations reach a constant amplitude because a stochastic Hill function is used. Another observation is that the introduction of extrinsic fluctuations causes a change in the period of the concentration of p53 and amplifies oscillations in the variance of p53, as shown in Fig. \ref{fig1}. 

\begin{figure}[h!t]
\begin{subfigure}{\linewidth}
\includegraphics[width=0.5\textwidth]{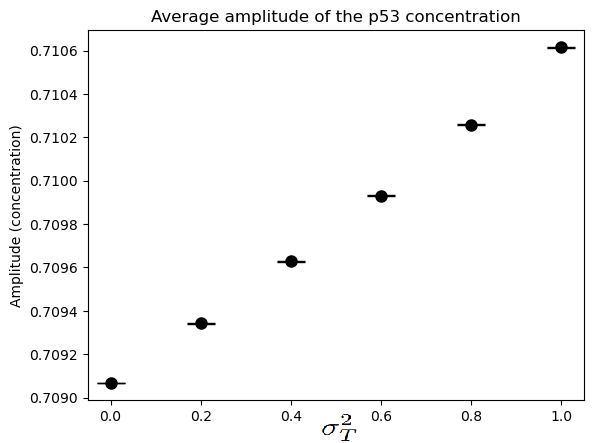}\hfill
\includegraphics[width=0.5\textwidth]{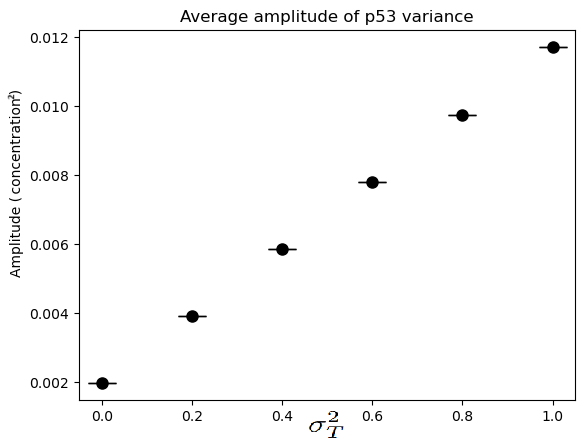}
\end{subfigure}
\begin{subfigure}{\linewidth}
\includegraphics[width=0.5\textwidth]{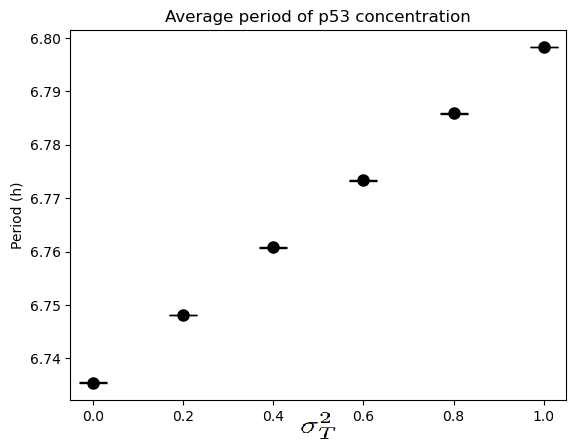}\hfill
\includegraphics[width=0.5\textwidth]{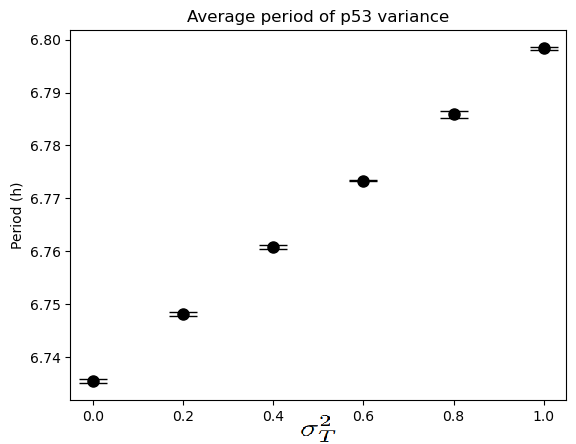}
\end{subfigure}
\caption{\textbf{Effect of stationary extrinsic fluctuations in the p53 dynamics.} We focus on the interval of 100-200 h, and then calculate the average of the amplitudes and periods for different values of $\sigma^2_T$. Left: Amplitude and period of p53 oscillations for increasing $\sigma^2_T$ (depicted by black circles). Right: Amplitude and period of p53 variance. A noticeable trend emerges across all panels: as $\sigma^2_T$ increases, both the average amplitude and the period of the system experience an upward trend. The substantial increase in the average amplitude of p53 variance is particularly striking. Furthermore, we observed the same increase in the average period of p53 concentration and the average period of p53 variance. The system was modeled using the parameters and initial conditions detailed in Table \ref{tabla2}.}
\label{fig2}
\end{figure}

We now quantify the effects of extrinsic fluctuations in the system, focusing on the amplitude and period of the oscillations, see Figure \ref{fig2}. This system has only one steady state in which the concentrations are positive; therefore, we used the same initial conditions as those used in Figure \ref{fig1} without loss of generality for this analysis. We focus on calculating the average amplitudes and periods for different values of $\sigma^2_T$ at interval of 100-200 h.

The increase in $\sigma^2_T$ is accompanied by a corresponding increase in the average amplitude of p53 oscillations. In particular, when $\sigma^2_T$ increases from zero to one, the average amplitude of p53 concentration experiences a modest increase of 0.22\%. Moreover, the average amplitude of the variance of p53 exhibited a notable increase with $\sigma^2_T$; specifically, we observed an increase of 490\% when moving from $\sigma^{2}_T=0$ to $\sigma^2_T=1$. Similarly, the average period of p53 concentration tended to increase with $\sigma^2_T$.  With the same increase in  $\sigma^2_T$, the average period showed an increase of approximately 0.93\%.
 This trend is mirrored in the average period of p53 variance, with an increase of approximately 0.063 when  $\sigma^2_T$ increases from zero to one, reflecting a comparable increase of 0.93 \%. Notably, these findings highlight a consistent pattern wherein both the average period of p53 concentration and the average period of p53 variance experienced similar increases.  

\subsection{Case II: Time-dependent extrinsic fluctuations}
\begin{figure}[h!t]
\begin{subfigure}{\linewidth}
\includegraphics[width=0.5\textwidth]{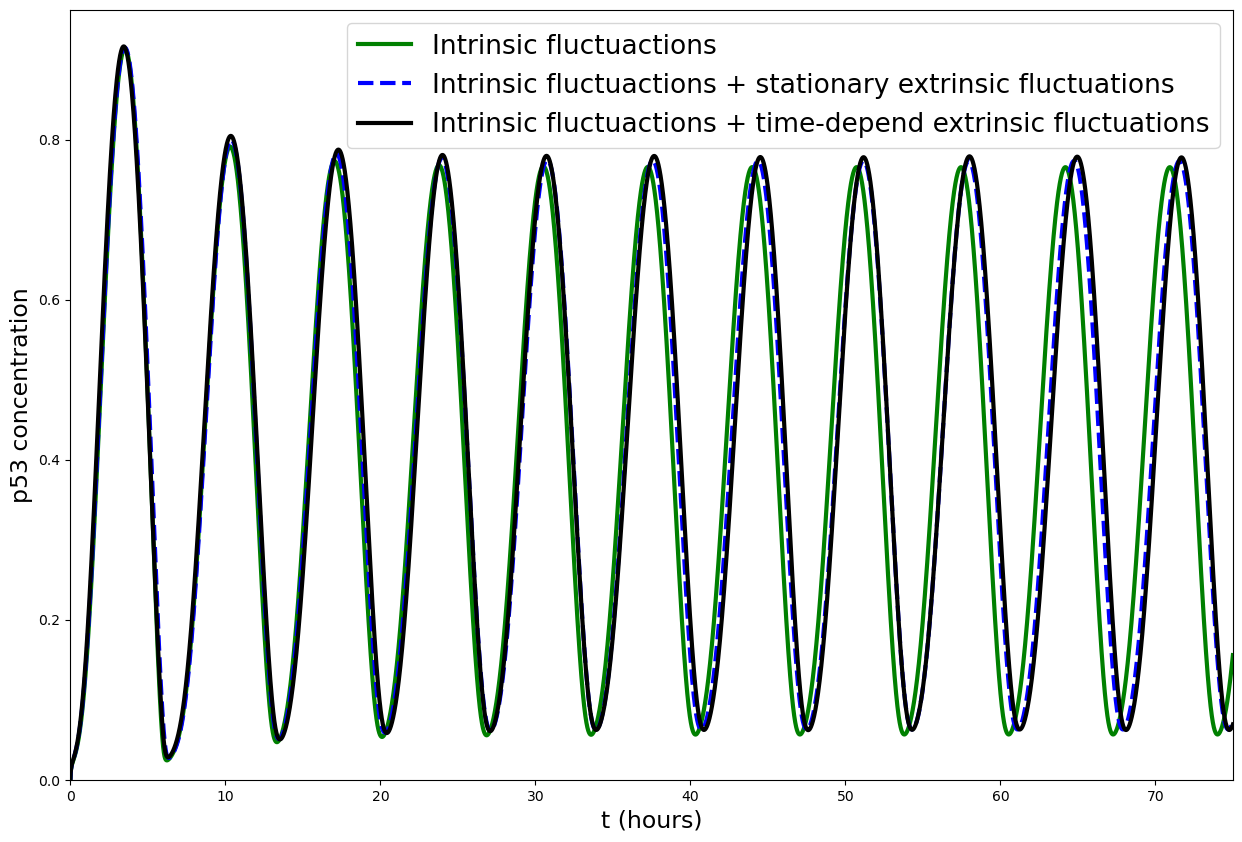}\hfill
\includegraphics[width=0.5\textwidth]{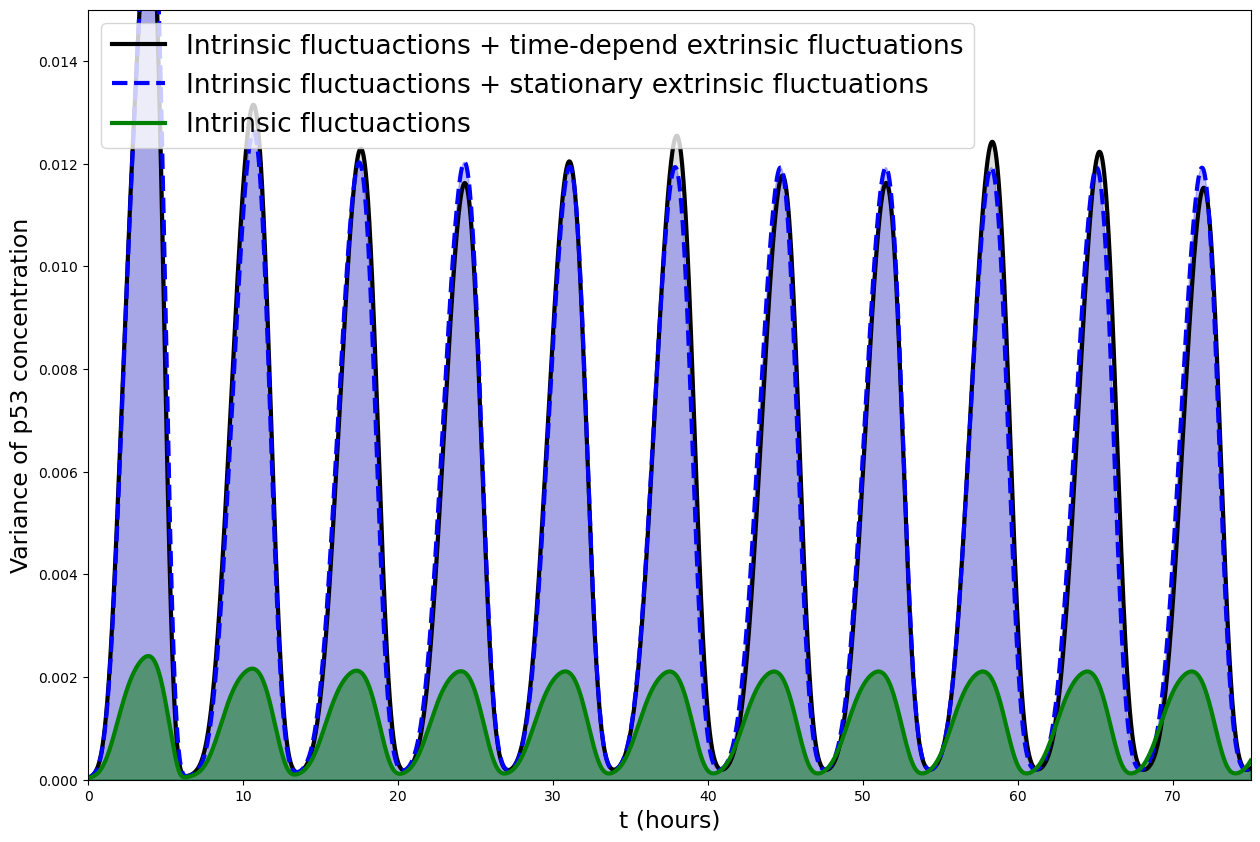}
\end{subfigure}
\caption{\textbf{Dynamics p53 concentration and its variance with TD-extrinsic fluctuations.} Dynamics of p53 concentration and its variance under the influence of intrinsic and TD-extrinsic fluctuations. We focus on the interval of 100-200 h, and then calculate the average of the amplitudes and periods for different values of $\sigma^2_T$.  The left figure depicts the dynamics of the involved species; green line: only intrinsic fluctuations; dashed blue line: intrinsic and stationary extrinsic fluctuations; and black line: intrinsic and TD-extrinsic fluctuations. When considering intrinsic and extrinsic fluctuations, either stationary and time-dependent, we observed an increase in the amplitude and phase shift. The figure on the right shows the variance of p53. Here, we note a similar trend; in the case where we include TD-extrinsic fluctuations alongside intrinsic fluctuations that result in an amplified amplitude and phase shift. For the case in which there are only intrinsic fluctuations, the system was modeled with the parameters and initial conditions listed in Table \ref{tabla1}. For the case in which there are intrinsic fluctuations and stationary extrinsic fluctuations, Table \ref{tabla2} and $\sigma^2_T=1$ are used. For the other case, Table \ref{tabla3} and $\sigma^2_T=1$ are used.}
\label{fig3}
\end{figure}

We extended the p53 model to incorporate time-dependent (TD) extrinsic fluctuations. To build this model, we used the Arrhenius equation again, but now we considered temperature oscillations over time, assuming that they mimic those observed in the human body \cite{Brown, Daka}, then we get,
\begin{align}
    k_i(t)=&  k^0_ie^{\frac{1}{40} \left( cos\left( \frac{\pi}{12} t  \right)\right)}, 
\end{align}
where $i,j=$ \{$1,2,3,4,5,6$\} and  $k_i^0$ denote the reaction constants evaluated at $T=309.65K$. 
To model the system we need to add a set of differential equations for $C^1_{l,j}$ ($l=$ \{$1,2,3$\}) and $C^2_{i,j}$, for this, we used (\ref{16}) and (\ref{17}), first we calculate $g_i(k_i)$ for this we taking the time derivative of $k_i(t)$,

\begin{align}
    g_i(k_i,t)=& - k_i(t) \frac{\pi}{480}sin\left( \frac{\pi}{12} t  \right),
\end{align}
we expect that when  $\frac{d C^2_{i,j} }{dt}=0$ Equation (\ref{20}) is recovered, then we propose
{\small
\begin{align}
    G_{ij}(\textbf{k},t) = \left(\frac{k_i(t)}{10}\right) \left(\frac{k_j(t)}{10}\right) \left( \frac{\sigma^2_T}{e^{\frac{1}{20} \left( cos\left( \frac{\pi}{12} t  \right)\right)} + \frac{\sigma^2_T}{100}} \right) \frac{\pi}{240} sin\left( \frac{\pi}{12} t  \right),
\end{align}}
$\sigma^2_T$ represents the variance in temperature and is considered to be constant, as described in the previous subsection. Then we get the next equations for $C^2_{i,j}$  (for more details, see Appendix \ref{A})
{\small
\begin{align}
    \frac{d C^2_{i,j} }{dt}=& \left( - C^2_{i,j} + \left( k_i(t) k_j(t) + C^2_{i,j} \right)  \left( \frac{\sigma^2_T}{100 e^{\frac{1}{20} \left( cos\left( \frac{\pi}{12} t  \right)\right)}+ {\sigma^2_T}} \right) \right)  \frac{\pi}{240} sin\left( \frac{\pi}{12} t  \right), \label{19}
\end{align}}
 
In Figure \ref{fig3}, we numerically solved the system for three cases: intrinsic fluctuations, intrinsic and stationary extrinsic fluctuations, and intrinsic and TD-extrinsic fluctuations.  We observed that oscillations were present even when TD-extrinsic fluctuations were introduced. However, p53 concentrations also exhibited a change in their period. Another observation is that the amplitude of the variance of p53 oscillates with respect to other cases because $k_i(t)$ oscillates.

\begin{figure}[h!t]
\begin{subfigure}{\linewidth}
\includegraphics[width=0.5\textwidth]{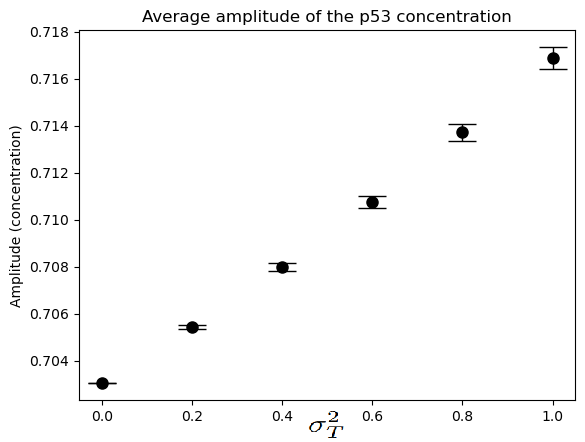}\hfill
\includegraphics[width=0.5\textwidth]{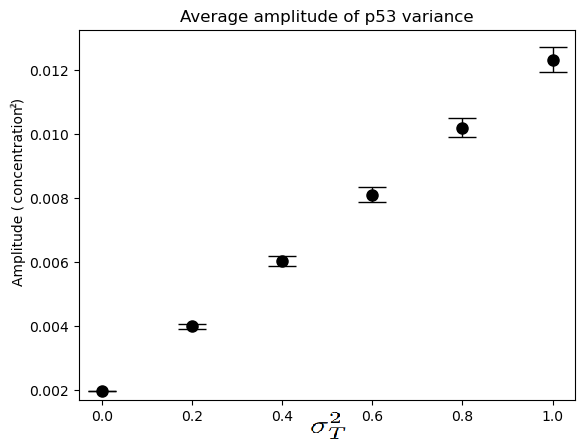}
\end{subfigure}
\begin{subfigure}{\linewidth}
\includegraphics[width=0.5\textwidth]{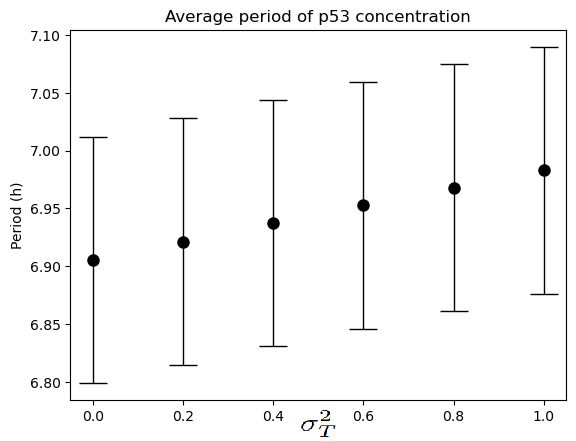}\hfill
\includegraphics[width=0.5\textwidth]{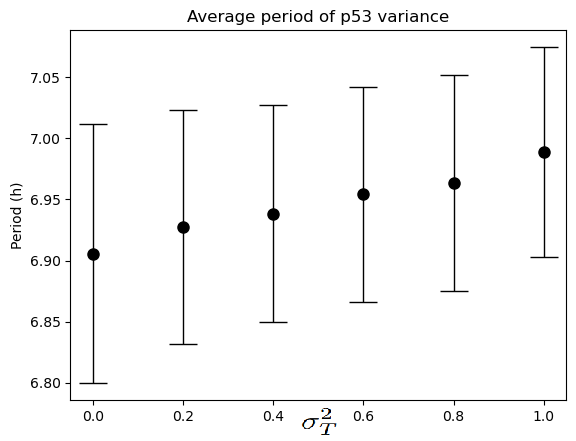}
\end{subfigure}
\caption{\textbf{Effect of TD-extrinsic fluctuations in p53 dynamics.} We focus on the interval of 100-200 h, and then calculate the average of the amplitudes and periods for different values of $\sigma^2_T$. In the figures on the left, we observe the impact of increasing $\sigma^2_T$ on the average amplitudes and periods of p53 (depicted by black circles). Additionally, we calculated the standard deviation of the averages (represented by vertical bars), although this appears to be relevant to the average of the period. On the right-hand side, we depict the average amplitude and period of variance of p53. A noticeable trend emerges across all panels: as $\sigma^2_T$ increases, both the average amplitude and the period of the system experience an upward trend.  Furthermore, we observed a considerable standard deviation in the average period of p53 concentration and the average period of p53 variance. The system was modeled using the parameters and initial conditions listed in Table \ref{tabla3}.}
\label{fig5}
\end{figure}

We focus on the interval of 100-200 h, and then calculate the average of the amplitudes and periods for p53 concentration and their variance, similarly to the previous section. Simulations were performed for different values of  $\sigma^2_T$, and the parameters used are listed in Table \ref{tabla3}.  The results of the simulations are shown in Figure \ref{fig5}, where we observe that as $\sigma^2_T$ increases, the amplitude of the p53 oscillations increases. When $\sigma^2_T$ increased from zero to one, the increment was approximately 1.97\%. Additionally, the amplitude of the variance followed a similar pattern, expanding as $\sigma^2_T$ increased, using the same increase in $\sigma^2_T$, which increases by 525 \%. Similarly, the p53 period tended to elongate with increasing $\sigma^2_T$; when $\sigma^2_T$ increased from zero to one, this elongation increased by 1.12 \%. Moreover, the variability in the variance of p53 increased when $\sigma^2_T$ increased from zero to one, reflecting an increase of approximately 1.12\%.

\begin{figure}[h!t]
 \centering
\begin{subfigure}{\linewidth}
\includegraphics[width=0.5\textwidth]{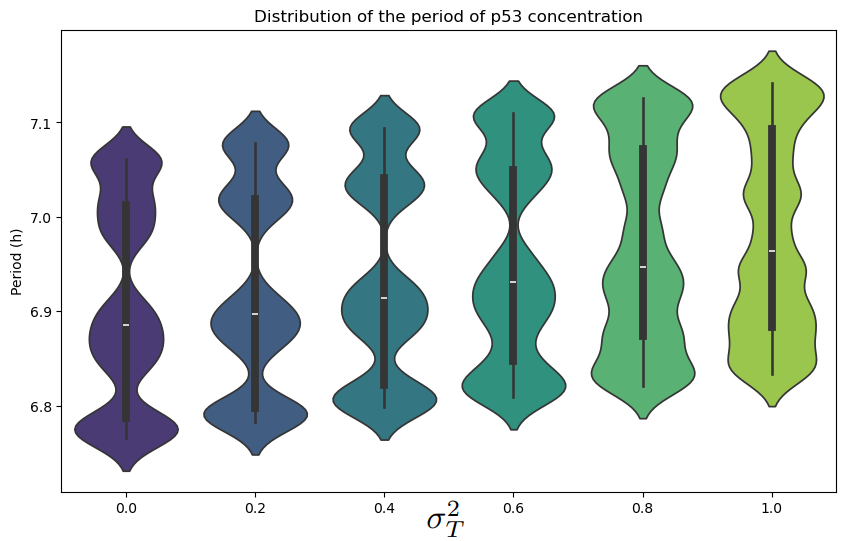}\hfill
\includegraphics[width=0.5\textwidth]{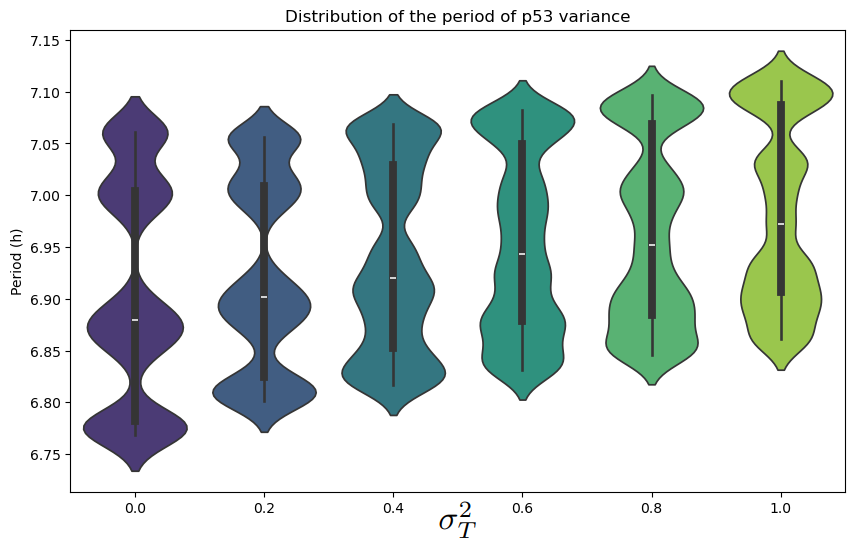}
\end{subfigure}
\caption{\textbf{Distribution of the period of concentration of p53 and its variance.}  We examined the distribution of the period of p53 concentration and variance for each $\sigma^2_T$, considering intrinsic fluctuations and TD-extrinsic fluctuations. We observe a growing trend in the period. Furthermore, as $\sigma^2_T$ increases, the period tends to be distributed more uniformly.  The system was modeled using the parameters and initial conditions listed in Table \ref{tabla3}.}
\label{fig6}
\end{figure}

A significant difference was observed compared with stationary extrinsic fluctuations. With time-dependent (TD) extrinsic fluctuations, the period of p53 concentration, as well as its variance, exhibits a considerable standard deviation. Figure \ref{fig6} shows the distributions of these periods, analyzing their distributions at interval of 100-200 h.  In this figure, we observe that at each value of $\sigma^2_T$, there is a distribution of periods and the center of the distribution increases as $\sigma^2_T$ increases. These distributions arise because the p53 concentration (and variance) exhibits oscillating periods owing to the oscillation of the variables $k_i(t)$.

\section{Results and Conclusions} \label{section5}


This study presents an approach based on a set of ordinary differential equations using the first two moments to examine extrinsic fluctuations in stochastic systems. This approach offers computational efficiency compared to the direct implementation of the chemical master equation, as the system is transformed into a set of ordinary differential equations.

The derived differential equations provide information about the dynamics of the system considering two types of fluctuations: intrinsic fluctuations, which are inversely proportional to the system size $\Omega$, and extrinsic fluctuations, which are equivalent to fluctuations in the parameters (rate constants). The accuracy of this approach is limited to systems with first-order reactions and/or small fluctuations ( intrinsic and extrinsic). For systems that do not have these characteristics, incorporating higher-order moments into the approximation can improve accuracy \cite{Lakatos}. Although this expansion increases the number of differential equations, the use of ordinary differential equations is computationally more efficient. 

Reproducing the results of previous research \cite{Vastola}, we applied this approach to analyze a biological oscillator, focusing on the p53 model and its response to temperature-induced extrinsic fluctuations. Although it is reasonable to expect drastic changes in system dynamics with the addition of extrinsic fluctuations, the oscillatory nature of the system persisted. 

Our findings underscore the impact of extrinsic fluctuations on the nature of oscillations in biological systems such as biological clocks. Notably, alterations in oscillatory behavior or dynamics depend on the characteristics of extrinsic fluctuations, emphasizing the importance of considering them in biological system modeling

In particular, we found that extrinsic fluctuations increased the oscillation amplitude, with the most significant change observed in the amplitude of p53 variance. This indicates that the concentration of p53 becomes more variable. Both the p53 concentration and variance increased when extrinsic fluctuations were introduced. Although the increase in each period was small, it could have a significant long-term impact,  and thus, two potential health consequences can arise. First, if the slowing of dynamics becomes significant, the suppression of DNA damage can be delayed, causing its accumulation. Second, the increased variability in p53 concentration could result in insufficient levels of p53.

Although the analyzed model is highly simplified, it highlights the factors that could contribute to cancer proliferation. 
Therefore, it is essential to study extrinsic fluctuations in more complex networks to analyze whether amplification or stabilization of the effects of extrinsic fluctuations occurs as the number of interactions in the system increases.
Such studies will enhance the development of alternative mechanisms for preventing unwanted scenarios related to health issues. 

Moreover, the proposed method of expanding moments around the mean holds promise for applications in diverse systems, as it allows the analysis of systems with many components and interactions by transforming the problem of solving the Master Chemical Equation into a problem of solving a system of ordinary differential equations. This is advantageous because computers do not have problems solving high-dimensional systems of ordinary differential equations while solving a high-dimensional master equation is not yet practical due to the memory needs.

\section*{Acknowledgments}
Manuel E. Hernández-García acknowledges the financial support of CONAHCYT through the program "Becas Nacionales 2023".\\
This work was supported by \textit{Programa de Apoyo a Proyectos de Investigación e Innovación Tecnológica} (PAPIIT UNAM; IA203524 to M.G.-S), and by ANID—Millennium Science Initiative Program—Millennium Institute for Integrative Biology (iBio; ICN17\_022 to M.G.-S).
Additionally, the authors want to thank for their support to Luis A. Aguilar from the \textit{Laboratorio Nacional de Visualización Científica Avanzada} and Jair Santiago García Sotelo, Iliana Martínez, Rebeca Mucino, and Eglee Lomelín from the \textit{Laboratorio Internacional de Investigación sobre el Genoma Humano, Universidad Nacional Autónoma de México, Santiago de Querétaro, México}. \\
Jorge Velázquez-Castro acknowledges Benemérita Universidad Autónoma de Puebla-VIEP financial support through project 00398-PV/2024.

\section*{Declarations}
The authors declare that there is no conflict of interest regarding the publication of this article. 

All data generated or analyzed during this study are included in this published article.

\appendix
\section{Exogenous Cox Ingersoll Ross model} \label{C}
The derived equations (\ref{10})–(\ref{17}) were applied to one of the systems presented in \cite{Vastola} for validation purposes. This is a model of RNA dynamics in which a function describes RNA synthesis in the system $K(t)$ (which is stochastic and varies with time) because it may be complicated to capture the transcription in detail. RNA subsequently matures and degrades as shown by the following reactions:  
\begin{align}
    0 \stackbin[]{K(t)}{\rightarrow}  N \stackbin[]{\beta}{\rightarrow} M \stackbin[]{\gamma}{\rightarrow} 0, \label{22.a}
\end{align}
where $N$ denotes nascent RNA, $M$ denotes mature RNA, $\beta$ is the synthesis rate, $\gamma$ is the degradation rate, and $K$($=K(t)$) (we omit to write explicit time dependence) is a function that follows the Cox–Ingersoll–Ross process. This is described by the following Langevin equation
\begin{align}
    \frac{d K}{dt}= k_0 -K + \sqrt{\frac{k_0 K}{a}} \epsilon(t), 
\end{align}
where $k_0$ is the deterministic constant rate, $a$ is a parameter that adjusts the intensity of the fluctuations, and $\epsilon(t)$ is a white Gaussian noise variable. Its corresponding Fokker-Plank equation is, 
{\footnotesize
\begin{align}
    \frac{d P(K,t)}{dt}= - \frac{\partial}{\partial K}[(k_0-  K)P(K,t)] + \frac{\partial^2}{\partial K^2}\left[\left( \frac{k_0 K}{a}\right) P(K,t) \right],  \label{B1}
\end{align}}
 First, we determine the matrix of the stoichiometric coefficients $\alpha_{ij}$ and $\beta_{ij}$,  and the stoichiometric matrix $\Gamma_{ij}$ from reactions (\ref{22.a}):

{\small
\begin{align}
 \alpha_{jl}&= \begin{pmatrix}
		0& 0  \\
		1& 0  \\
		0& 1   
	\end{pmatrix} ,    &
 \beta_{jl}&=  \begin{pmatrix}
		1& 0  \\
		0& 1  \\
		0& 0  
	\end{pmatrix},  &
 \Gamma_{lj}&= \begin{pmatrix}
		1& -1 & 0  \\
		0& 1  & -1
	\end{pmatrix}, \nonumber
\end{align}}
using this, we calculated the reaction rates of the system as follows:
\begin{align}
   R_1^{D}&= k,  \nonumber \\
   R_2^{D}&= \beta n , \nonumber \\
    R_3^{D}&= \gamma m  , \label{B2}
\end{align}
where $n$ and $m$ represent the concentrations of $N$ and $M$, respectively, and $k$($=k(t)$) is the mean of $K(t)$, we omit to write explicit time dependence.  From (\ref{B1}), we find that $g(k)=k_0 - k$ and $G(k)=\frac{2 k_0 k}{a}$. Using these elements and following the procedure developed in Section \ref{section 3}, we derive a set of differential equations that describe the system. We have the following equations for $n$, $m$ and $k$
\begin{align}
    \frac{d n}{dt}=& k - \beta n, \nonumber \\
    \frac{d m}{dt}=& \beta n - \gamma m, \nonumber \\
    \frac{d k}{dt}=& k_0 - k.
\end{align}
We now derive the covariances and correlations between system variables,
{\small
\begin{align}
    \frac{d \sigma^2_{n,n}}{dt}=& \frac{\beta n}{\Omega} - 2 \beta \sigma^2_{n,n}  + \frac{k}{\Omega} + 2 C^1_{n,k}, \nonumber \\
    \frac{d \sigma^2_{m,m}}{dt}=& \frac{\beta n + \gamma m}{\Omega} + 2 \beta \sigma^2_{m,n} - 2 \gamma \sigma^2_{m,m}, \nonumber \\
    \frac{d \sigma^2_{n,m}}{dt}=& -\frac{\beta n}{\Omega} - (\beta + \gamma) \sigma^2_{n,m} + \beta \sigma^2_{n,n} + C^1_{m,k}, \nonumber  \\
    \frac{d C^1_{n,k}}{dt}=& -\beta C^1_{n,k} + C^2_{k,k} -  C^1_{n,k}, \nonumber  \\
    \frac{d C^1_{m,k}}{dt}=& \beta C^1_{n,k} - \gamma C^1_{m,k} - C^1_{m,k}, \nonumber  \\
     \frac{d C^2_{k,k}}{dt}=& -2  C^2_{k,k} + \frac{2 k_0 k}{a}.
 \end{align}}

These results are similar to those derived by \cite{Vastola}. However, in this case, we directly substitute (\ref{B2}), $g(k)$ and $G(k)$ into the set of differential equations derived in section \ref{section 3}, thereby allowing for a more straightforward derivation. The reactions in this model are first-order, then we have an exact description of the system. Next, we simulate the differential equations of the proposed system.

\begin{figure}[h!t]
\begin{subfigure}{\linewidth}
\includegraphics[width=0.5\textwidth]{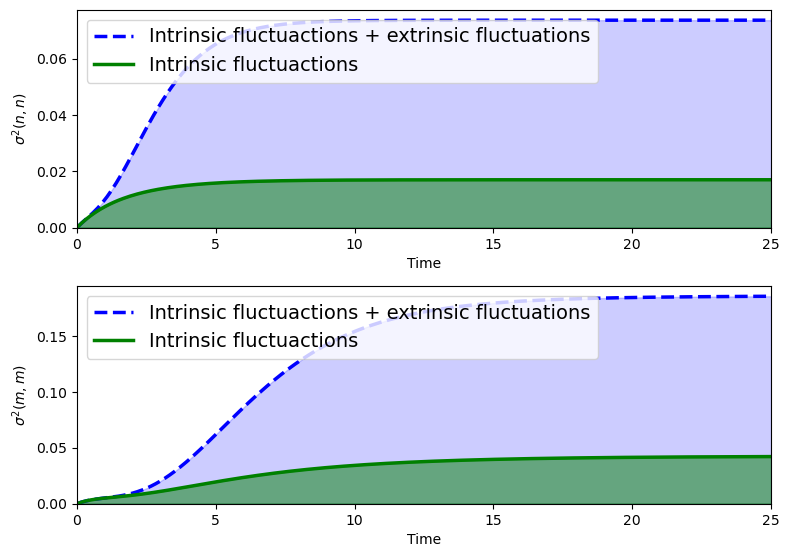}\hfill
\includegraphics[width=0.5\textwidth]{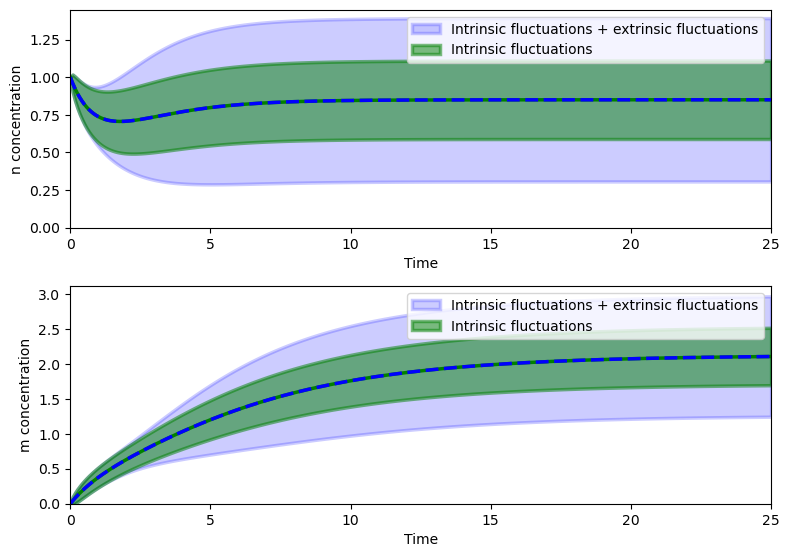}
\end{subfigure}
\caption{ \textbf{Simulation transcriptional model with Cox Ingersoll Ross model.} Dynamics of the species involved and their variance. We modeled the system using only intrinsic fluctuations (green). When extrinsic fluctuations are introduced (purple), the size of the fluctuations increases significantly. The figure on the right side shows the dynamics of the species involved. The filled areas represent the spaces where intrinsic fluctuations are (green) and where both intrinsic and extrinsic fluctuations are (purple). The system was modeled using the following parameters: $\beta=0.5 h^{-1}$, $\gamma=0.2 h^{-1}$, $a=4.25$,  $k_0=0.425$ $mol h^{-1}$, and $\Omega=500$ (molecules $mol^{-1}$). The initial conditions were: $n(0)=1$ $mol$, $m(0)=0$ $mol$, and $k(0)=0$ $mol h^{-1}$.}
\label{fig.B}
\end{figure}

Figure \ref{fig.B} shows the effects of the extrinsic fluctuations on the system. The left panel shows the variance of the system variables, indicating a significant increase in the size of fluctuations when both intrinsic and extrinsic fluctuations are considered (purple) compared with when only intrinsic fluctuations are present (green). The panel on the right shows the averages of the involved species and the range of values where fluctuations are observed, with only intrinsic (green) and intrinsic and extrinsic (purple) fluctuations demonstrating a substantial increase in the fluctuation size when extrinsic fluctuations are also present.

\section{Size-dependent model of p53} \label{D}
\begin{figure}[h!t]
\begin{subfigure}{\linewidth}
\includegraphics[width=0.5\textwidth]{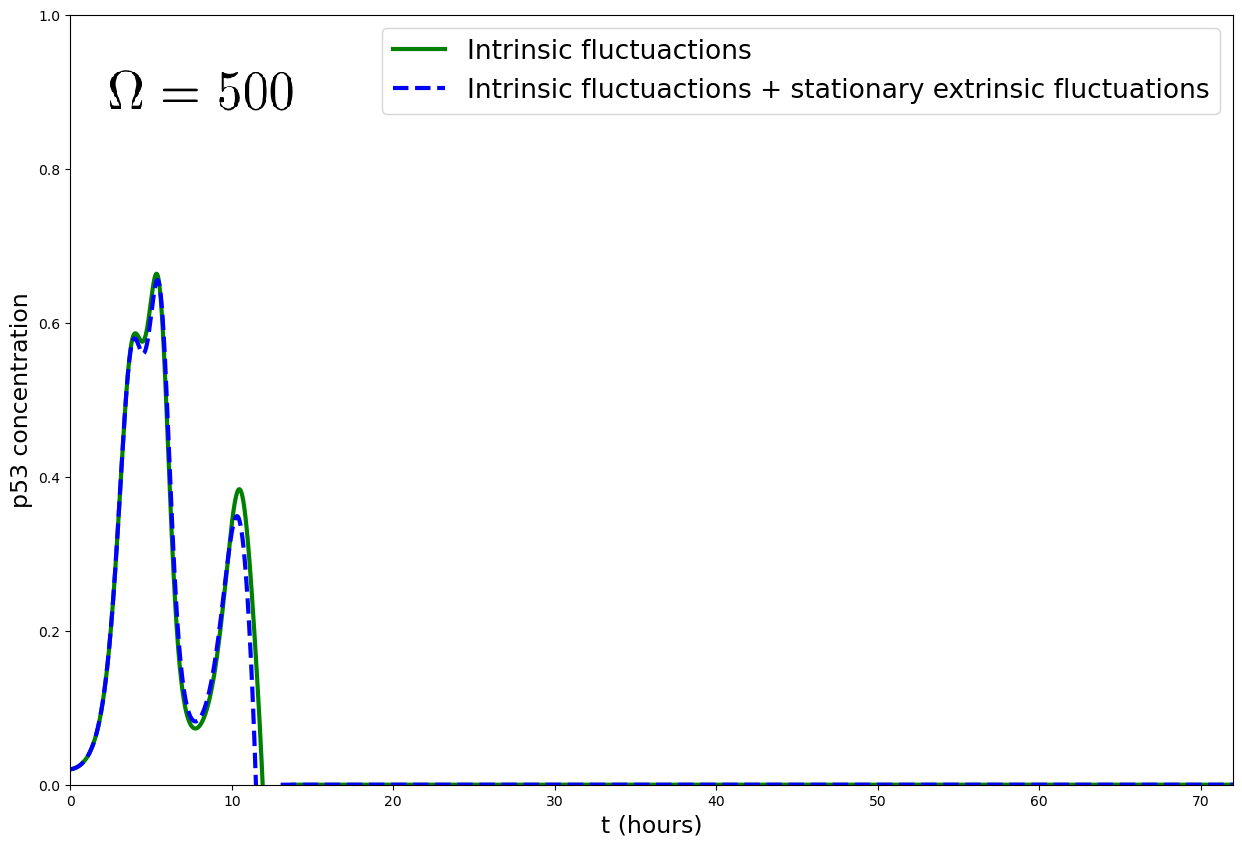}\hfill
\includegraphics[width=0.5\textwidth]{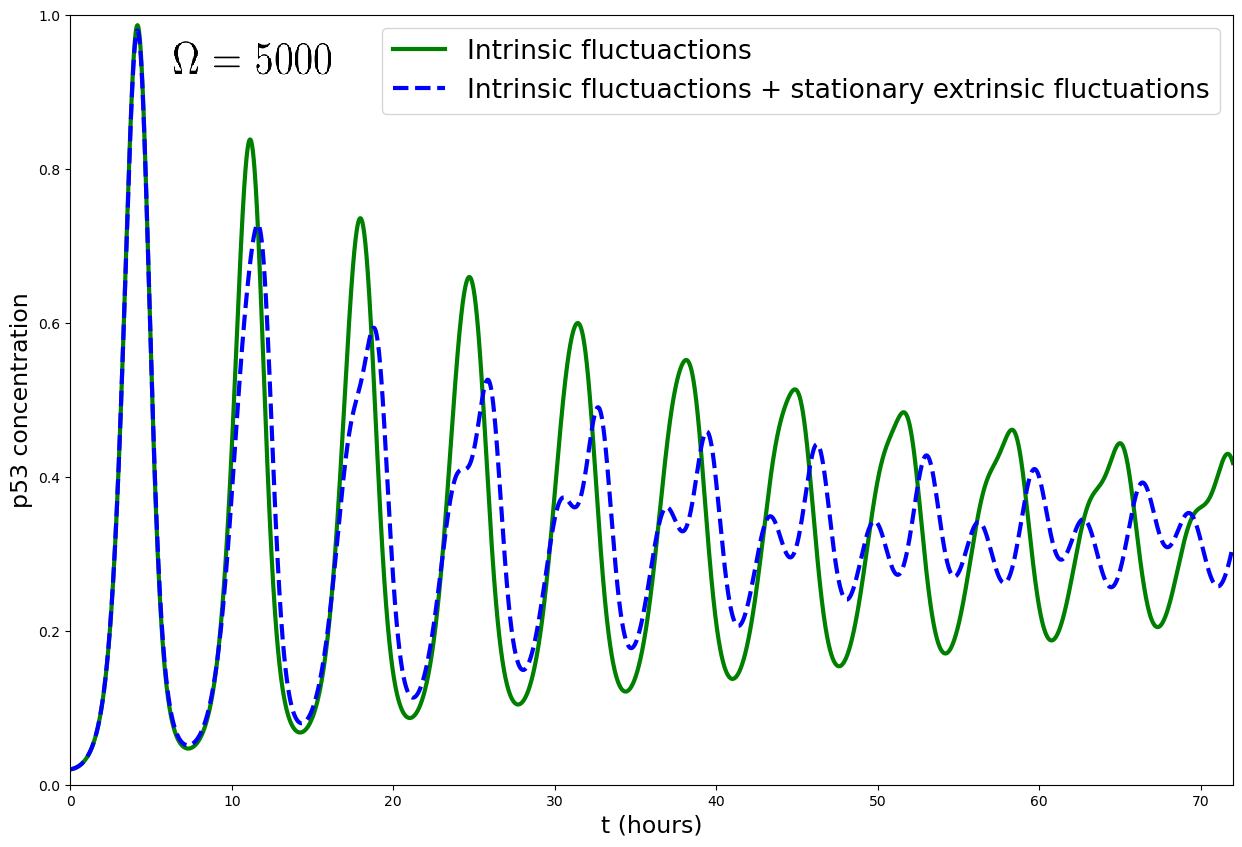}
\end{subfigure}
\caption{ \textbf{Size-dependent model of p53.}  In this figure, we observe the dynamics of p53 when considering only intrinsic fluctuations (green line) and both intrinsic and stationary extrinsic fluctuations (dashed blue lines). We observed that when the size of the system is $\Omega=500$ (molecules $mol^{-1}$), p53 concentration goes to zero, both considering only intrinsic fluctuations or intrinsic and extrinsic fluctuations. However, if the size of the system is $\Omega=5000$ (molecules $mol^{-1}$), damped oscillations appear, approaching a steady state distinct from zero. The system was modeled using the following parameters: $k_1=2$ $h^{-1}$, $k_2=3.7$ $mol^{-1} h^{-1}$, $k_3=1.5$ $h^{-1}$, $k_4=1.1$ $h^{-1}$, $k_5=0.9$ $h^{-1}$ and $\sigma^2_T=0.05$. The initial conditions were: $x_1(0)=0.02$ $mol$, $x_2(0)=0.2$  $mol$, and $x_3(0)=0.5$ $mol$.}
\label{figd}
\end{figure}

We also analyzed an alternative model presented in \cite{Zato}. However, upon conducting a system analysis using our differential equation framework, we observed that if we choose the parameter set outlined in \cite{Zato}, the stability of the system dynamics becomes size-dependent, as we will see next. For example, when $\Omega=500$ (molecules $mol^{-1}$), the concentration of p53 approached zero, as shown in Figure \ref{figd}. 

In this system, we have the following reactions.

\begin{table}[h!]
    \centering
    \begin{tabular}{|c|c|}
        $\emptyset  \stackbin{k_1}{\longrightarrow} P53$ & $P53 + \text{Mdm2}  \stackbin{k_2}{\longrightarrow} \text{Mdm2}$ \\
          $P53  \stackbin{k_3}{\longrightarrow} P53 + \text{Mdm2 precursor}$ & $\text{Mdm2 precursor}  \stackbin{k_4}{\longrightarrow} \text{Mdm2}$ \\
          $\text{Mdm2} \stackbin{k_5}{\longrightarrow} \emptyset$ & \\
    \end{tabular}
\end{table}

We have the next stoichiometric matrix
{\small
\begin{align}
 \Gamma_{lj}&= \begin{pmatrix}
		1& -1 &  0 & 0 &0 \\
		0& 0  & 1 & -1 &0 \\
        0& 0  & 0 & 1  &-1
	\end{pmatrix}, \nonumber
\end{align}}
the reaction rates of the system,
\begin{align}
   R_1^{D}&= k_1 x_1, &
    R_2^{D}= k_2 x_1x_3, \nonumber \\
   R_3^{D}&= k_3 x_1 , &
    R_4^{D}= k_4 x_2, \nonumber \\
    R_5^{D}&= k_5 x_3, &
     \label{21}
\end{align}
where $x_1$, $x_2$, and $x_3$ represent the concentrations of p53, Mdm2 precursor, and Mdm2, respectively. $k_1$ is the p53 synthesis rate, $k_2$ is the p53 degradation rate by Mdm2, $k_3$ is the p53-dependent Mdm2 production rate, $k_4$ is the Mdm2 maturation rate, and $k_5$ is the Mdm2 degradation rate.  

We conducted a modeling exercise and the results are shown in Figure \ref{figd}. This figure distinctly illustrates the disparities when considering intrinsic fluctuations alone versus when incorporating both intrinsic and stationary extrinsic fluctuations. From this figure, we can see that the concentration of p53 approaches zero when $\Omega=500$ (molecules $mol^{-1}$). In contrast, when $\Omega=5000$ (molecules $mol^{-1}$), the concentrations of the molecules exhibit damped oscillations, eventually reaching a steady state distinct from zero. This behavior persists even when extrinsic fluctuations are included, indicating that $\Omega$ is the principal factor that influences the stability of the system. Consequently, this model is unsuitable for analyzing systems with a low number of molecules. It is important to study the phase space to understand the dynamics of this system thoroughly. However, we did not perform this analysis because it was beyond the scope of this study.

\section{Fluctuations by temperature in reaction constants } \label{A}
\subsection{Case I: Stationary}
To introduce fluctuations in the reaction rate constants of the system, we assumed that reaction rates could be affected by the temperature. To do this, we will assume that they follow the Arrhenius equation \cite{Atkin},
\begin{align}
    k(T)=A e^{-\frac{E_a}{RT}}, \label{a1}
\end{align}
where $k$ is the reaction rate constant, $A$ is the frequency factor (collision frequency between molecules), $E_a$ is the activation energy, $R$ is the ideal gas constant, and $T$ is the temperature in Kelvins. Therefore, if we assume that the temperature fluctuates, then Equation (\ref{a1}) becomes
\begin{align}
    k(T + \delta T)=A e^{-\frac{E_a}{R(T+\delta T)}}, 
\end{align}
where $\delta T$ denotes the temperature fluctuation. Assuming the fluctuation is small, we can perform a first-order expansion
\begin{align}
    k(T + \delta T)\approx& A e^{-\frac{E_a}{RT}}\left( 1 + \frac{E_a}{R T^2}\delta T \right)  \nonumber \\
    =&k(T) \left( 1 + \frac{E_a}{R T^2}\delta T \right) \nonumber \\
    =& k(T) + \eta_{k(T)}, 
\end{align}
($\eta_{k(T)}=k(T) \left( \frac{E_a}{R T^2}\delta T \right)$), fluctuations can be introduced into the system, considering that the temperature fluctuates. From this same equation, we will calculate the covariances of $k$, so we have
\begin{align}
    C^2_{k,k}= \braket{(\eta_{k(T)})(\eta_{k(T)})} = \left(  \frac{E_a}{R T^2} k \right)^2 \sigma^2_T.
\end{align}
($\sigma^2_T= \braket{(\delta T)(\delta T)}$) With this formula, our system depends only on the covariance of temperature. Assuming that the temperature at which these experiments are conducted is around $309.65 K$, which is approximately body temperature, as well as $R=8.314 \frac{J}{mol K}$, and taking $E_a=80 \frac{kJ}{mol}$ as the approximate activation energy for protein synthesis and degradation  ($E_a$ is a different quantity for each particular system) \cite{Voet}, now the covariance of the reaction rates will be finally
\begin{align}
    C^2_{k,k} \approx \left(\frac{k}{10}\right)^2 \sigma^2_T. \label{a5}
\end{align}
Thus, the covariances of the reaction rate constants depend only on the temperature variance.

\subsection{Case II: Time-depend}
Temperature can be time-dependent; for example, in the human body, the temperature fluctuates. To describe this, we rely on \cite{Brown} to provide a formula for the temperature behavior in the human body, and also we considered the results in \cite{Daka} for give the amplitude of the oscillations, then we get 
\begin{align}
    T(t)= T_{max} + \frac{1}{4} cos\left( \frac{\pi}{12} t  \right), \label{a8}
\end{align}
where  $T(t)$ is the temperature in Kelvin and  $T_{max}=309.65 K$ indicating that the body temperature oscillates around it within a period of 24 h, and we have an amplitude of $0.25 K$. Now we substitute (\ref{a8}) into (\ref{a1}), from which we obtain,
\begin{align}
    k(t)=A e^{ \left( {-\frac{E_a}{R\left( T_{max} + \frac{1}{4} cos\left( \frac{\pi}{12} t  \right) \right)}} \right)}, \label{28}
\end{align}
we redefine this expression because this quantity oscillates around a value, which is $k^0= A e^{-\frac{E_a}{RT_{max}}}$, with this condition we get,
\begin{align}
    k(t) \approx k^0 e^{ \left(\frac{1}{40}  cos\left( \frac{\pi}{12} t  \right) \right)}. 
\end{align}
In this way, it will be easier to employ this function, and now all our quantities depend solely on time. We differentiate this function with respect to time,
\begin{align}
    \frac{d k(t)}{dt}= -k(t) \frac{\pi}{480} sin\left( \frac{\pi}{12} t  \right), 
\end{align}
with this result and from section \ref{section 3} we can find that 

\begin{align}
    g(k,t)= -k(t) \frac{\pi}{480} sin\left( \frac{\pi}{12} t  \right).
\end{align}
For modeling our system we need the next equation
\begin{align}
\frac{d C^2_{k,k} }{dt}=& 2 C^2_{k,k} \left( \frac{\partial g(k,t) }{ \partial k}  \right)+ \left( G(k, t) + \frac{C^2_{k,k}}{2} \frac{d^2 G(k, t) }{ d k^{ 2}} \right) ,     \nonumber
\end{align}
we expect that, when $\frac{d C^2_{k,k} }{dt}=0$ Equation (\ref{a5}) is recovered. Therefore, we propose
{\footnotesize
\begin{align}
    G(k,t)= \left(\frac{k(t)}{10}\right)^2 \left( \frac{\sigma^2_T}{e^{ \left(\frac{1}{20}  cos\left( \frac{\pi}{12} t  \right) \right)}+ \frac{\sigma^2_T}{100}} \right) \frac{\pi}{240} sin\left( \frac{\pi}{12} t  \right),
\end{align}}
$\sigma^2_T$ is a constant that represents the value of the variance of the temperature and is the same that appears in Equation (\ref{a5}), with this,  we get  
\begin{align}
\frac{d C^2_{k,k} }{dt}=& \left( - C^2_{k,k} + \left( k^2(t) + {C^2_{k,k}} \right) \left( \frac{\sigma^2_T}{100e^{ \left(\frac{1}{20}  cos\left( \frac{\pi}{12} t  \right) \right)}+ {\sigma^2_T}} \right) \right)  \frac{\pi}{240} sin\left( \frac{\pi}{12} t  \right) .    \label{a14}
\end{align}
Now, we can model our system considering that the variable $k$ varies and fluctuates over time. 

\section{Parameters and initial conditions}
Tables \ref{tabla1}, \ref{tabla2}, and \ref{tabla3} show the parameters and initial conditions used in each case: intrinsic fluctuations, intrinsic fluctuations and stationary extrinsic fluctuations, and intrinsic fluctuations and time-dependent extrinsic fluctuations, respectively.

\begin{table}[htbp]
\centering
\begin{tabular}{|p{3cm}|c|c|c|}
\hline
\textbf{Parameters and initial conditions} & \textbf{Description} & \textbf{Value} \\
\hline
$k_1$ & p53 synthesis rate. & 0.35 $h^{-1}$ \\
$k_2$ & p53 degratation rate. & 0.001 $h^{-1}$ \\
$k_3$ & Saturated p53 degradation rate. & 0.2 $h^{-1}$ \\
$k_4$ & p53-dependent Mdm2 production rate. & 0.55 $h^{-1}$ \\
$k_5$ & Mdm2 maturation rate.  & 0.2 $h^{-1}$ \\
$k_6$ &  Mdm2 degradation rate. & 0.25 $h^{-1}$ \\
$A_1$ & p53 threshold for degradation by Mdm2. & 0.01 $mol$ \\
$\Omega$ & Size of the system. & 500 (molecules) \\
$x_1(0)$ & Initial concentration of p53. & 0 $mol$ \\
$x_2(0)$ & Initial concentration of Mdm2 precusor. & 0.1 $mol$ \\
$x_3(0)$ & Initial concentration of Mdm2. & 0.8  $mol$ \\
$\sigma^2(x_i,x_j) (0)$ & Initial covariance of species. & 0 $mol^2$ \\
\hline
\end{tabular}
\caption{\textbf{Parameters and initial conditions for the model with only intrinsic fluctuations.} ($i=$ \{$1,2,3,4,5,6$\}) In this table, we show the parameters used for the system in the case in which there are only intrinsic fluctuations. These parameters and the initial conditions of the concentrations were obtained from \cite{Zato}.}
\label{tabla1}
\end{table}

\begin{table}[htbp]
\centering
\begin{tabular}{|p{3cm}|c|c|c|}
\hline
\textbf{Parameters and initial conditions} & \textbf{Description} & \textbf{Value} \\
\hline
$k_1$ & p53 synthesis rate. & 0.35 $h^{-1}$ \\
$k_2$ & p53 degratation rate. & 0.001 $h^{-1}$ \\
$k_3$ & Saturated p53 degradation rate. & 0.2 $h^{-1}$ \\
$k_4$ & p53-dependent Mdm2 production rate. & 0.55 $h^{-1}$\\
$k_5$ & Mdm2 maturation rate.  & 0.2 $h^{-1}$ \\
$k_6$ &  Mdm2 degradation rate. & 0.25 $h^{-1}$ \\
$A_1$ & p53 threshold for degradation by Mdm2. & 0.01 $mol$  \\
$\Omega$ & Size of the system. & 500 (molecules) \\
$x_1(0)$ & Initial concentration of p53. & 0 $mol$ \\
$x_2(0)$ & Initial concentration of Mdm2 precusor. & 0.1 $mol$ \\
$x_3(0)$ & Initial concentration of Mdm2. & 0.8 $mol$ \\
$\sigma^2(x_i,x_j) (0)$ & Initial covariance of species. & 0  $mol^2$ \\
$C^1(x_i,k_j) (0)$ & Initial covariance between species and parameters. & 0 $mol$ $h^{-1}$ \\
$C^2(k_i,k_j) $ & Covariance of parameters. & $\left(\frac{k_i}{10}\right)\left(\frac{k_j}{10}\right) \sigma^2_T$  \\
\hline
\end{tabular}
\caption{\textbf{Parameters and initial conditions for the model with intrinsic fluctuations and stationary extrinsic fluctuations.} ($i=$ \{$1,2,3,4,5,6$\}) In this table, we show the parameters used for the system in the case in which there are intrinsic fluctuations and stationary extrinsic fluctuations. These parameters and the initial conditions of the concentrations were obtained from \cite{Zato}. The covariance between the parameters is derived in Appendix \ref{A} and $\sigma^2_T$ is an adimensionalized constant parameter that we varied in our analysis.}
\label{tabla2}
\end{table}

\begin{table}[htbp]
\centering
\begin{tabular}{|p{3cm}|c|c|c|}
\hline
\textbf{Parameters and initial conditions} & \textbf{Description} & \textbf{Value} \\
\hline
$k_1^0$ & p53 synthesis rate at $T=309.65 K$. & 0.35 $h^{-1}$ \\
$k_2^0$ & p53 degratation rate at $T=309.65 K$. & 0.001 $h^{-1}$ \\
$k_3^0$ & Saturated p53 degradation rate at $T=309.65 K$. & 0.2  $h^{-1}$\\
$k_4^0$ & p53-dependent Mdm2 production rate at $T=309.65 K$. & 0.55 $h^{-1}$\\
$k_5^0$ & Mdm2 maturation rate at $T=309.65 K$.  & 0.2 $h^{-1}$ \\
$k_6^0$ &  Mdm2 degradation rate at $T=309.65 K$. & 0.25 $h^{-1}$ \\
$A_1$ & p53 threshold for degradation by Mdm2. & 0.01 $mol$  \\
$\Omega$ & Size of the system. & 500 (molecules) \\
$x_1(0)$ & Initial concentration of p53. & 0 $mol$ \\
$x_2(0)$ & Initial concentration of Mdm2 precusor. & 0.1 $mol$ \\
$x_3(0)$ & Initial concentration of Mdm2. & 0.8 $mol$ \\
$\sigma^2(x_i,x_j) (0)$ & Initial covariance of species. & 0 $mol^2$ \\
$C^1(x_i,k_j) (0)$ & Initial covariance between species and parameters. & 0 mol $h^{-1}$ \\
$C^2(k_i,k_j)(0) $ & Initial covariance of parameters. & $\left(\frac{k_i^0}{10}\right)\left(\frac{k_j^0}{10}\right) \sigma^2_T$  \\
\hline
\end{tabular}
\caption{\textbf{Parameters and initial conditions for the model with intrinsic fluctuations and time-dependent extrinsic fluctuations.} ($i=$ \{$1,2,3,4,5,6$\}) In this table, we show the parameters used for the system in the case in which there are intrinsic fluctuations and time-dependent extrinsic fluctuations. These parameters and the initial concentration conditions were obtained and adapted from \cite{Zato}. To expedite the convergence, the covariance between the parameters adopts a structure similar to that in Table \ref{tabla2}. Notably, this facilitated the rapid attainment of oscillatory values. $\sigma^2_T$ represents an adimensionalized constant parameter subject to variation in our analysis.}
\label{tabla3}
\end{table}

\end{document}